\documentclass[aps,prl,superscriptaddress]{revtex4-2}

\usepackage{amsmath,amssymb}
\usepackage{amsfonts}
\usepackage{graphicx} % Required for inserting images
\usepackage[a4paper, top=0.75in, bottom=1in, left=0.75in, right=0.75in]{geometry} % Set custom margins individually
\usepackage{mathrsfs}
\usepackage{verbatim}
\usepackage{pgfplots}
\pgfplotsset{compat=1.17}
\usepackage{blochsphere}
\usetikzlibrary{positioning,arrows,calc,math,angles,quotes}
\usepackage{blochsphere}
\usepackage{etoolbox}

\usepackage{braket}
\usepackage{comment}
\usepackage{mathtools}
\usepackage{tikz}
\usetikzlibrary{matrix, arrows.meta}

\usepackage{amsthm}

\newtheorem{theorem}{Theorem}[section] % Numbered by sections
\newtheorem{lemma}[theorem]{Lemma}     % Shares numbering with Theorem
\newtheorem{proposition}[theorem]{Proposition}
\newtheorem{corollary}[theorem]{Corollary}

\theoremstyle{definition}
\newtheorem*{definition}{Definition}
\theoremstyle{remark}
\newtheorem*{remark}{Remark}

\begin{document}

\title{The State-Operator Clifford Compatibility: A Real Algebraic Framework for Quantum Information}
\author{Kagwe A. Muchane}
\affiliation{Department of Computer Science, North Carolina State University}
\affiliation{Department of Mathematics, North Carolina State University}

\begin{abstract}
We revisit the Pauli--Clifford connection to introduce a real, grade-preserving algebraic framework for $n$-qubit quantum computation based on the tensor product $\mathcal{C}\ell_{2,0}(\mathbb{R})^{\otimes n}$. In this setting, the bivector $J = e_{12}$ satisfies $J^{2} = -1$ and supplies the complex structure on the $J$-closure of a minimal left ideal via right multiplication, while Pauli operations arise as left actions of Clifford elements. The Peirce decomposition organizes the algebra into sector blocks determined by primitive idempotents, with nilpotent elements generating transitions between sectors. Quantum states are represented as equivalence classes modulo the left annihilator, exhibiting the quotient description underlying the minimal left ideal. Adopting a canonical stabilizer mapping, the $n$-qubit computational basis state $\ket{0\cdots 0}$ is given natively by a tensor product of these idempotents. This structural choice leads to a compatibility law that is stable under the geometric product for $n$ qubits and aligns symbolic Clifford multiplication with unitary evolution on the Hilbert space.
\end{abstract}

\maketitle

\pagestyle{plain}

\section{Introduction}

Dimensions that are powers of two occupy a privileged position in quantum information, underlying both qubit Hilbert spaces and the representation theory of Clifford (geometric) algebras. The Clifford algebra has long been used to describe spin, relativistic field theory, and spinors in a coordinate-free language, most famously by Paul Dirac himself \cite{dirac1928}, and rediscovered in Hestenes' spacetime algebra \cite{Hestenes2015} and Lounesto's classification of spinor fields \cite{Lounesto2001}. Despite this fundamental correspondence, the application of geometric algebra to $n$-qubit quantum information science has remained fragmented, due in part to its historically niche and occasionally controversial standing within mathematics and physics.

Within quantum theory, the Pauli matrices have traditionally been recognized as a matrix representation of the Clifford algebra $\mathcal{C}\ell_{3,0}(\mathbb{R})$. Less frequently emphasized is that this mapping introduces representational redundancy, which persists under complexification. More recent theoretical work \cite{Silva2025UnifiedClifford, Hrdina2022CliffordQC} and general Clifford analysis confirm the isomorphism between the complex Clifford algebra $\mathcal{C}\ell_{2n,0}(\mathbb{C})$ and the $n$-qubit operator space $M_{2^n}(\mathbb{C})$. These approaches typically employ global constructions of the $2n$-dimensional algebra, often using Witt bases or non-canonical identifications such as $e_1 \leftrightarrow \sigma_x$ and $e_2 \leftrightarrow \sigma_y$. While convenient in fermionic contexts, such methods rely on a commuting pseudoscalar and a globally complexified algebra that obscures locality and grade decomposition. For computational quantum information science, this introduces two key limitations: it separates operator action from the algebraic geometry of states, and obscures the minimal real degrees of freedom underlying stabilizer dynamics.

This work establishes an alternative, compact framework based on the graded tensor product $\mathcal{C}\ell_{2,0}(\mathbb{R})^{\otimes n}$, and we show that the full complex behavior of a single qubit already appears within the four-dimensional real Clifford algebra, where the bivector $J=e_{12}$ supplies the complex structure via right multiplication on a minimal left ideal. Crucially, we introduce the canonical stabilizer mapping that aligns the algebraic generators directly with the generators of the $n$-qubit Pauli group, which allows the $n$-qubit vacuum state $|0\dots0\rangle$ to be defined natively by the tensor product of real algebraic idempotents, preserving grade and circumventing the complex machinery of the Witt basis. The key insight is the \textit{State-Operator Compatibility} under the geometric product,
\[
U(A P_n) = (U A) P_n,
\]
which holds for any Clifford element $U$ and any multivector $A$ in $\mathcal{C}\ell_{2,0}(\mathbb{R})^{\otimes n}$, with $P_n$ the $n$-qubit vacuum idempotent. This principle expresses the compatibility of left-acting operators and right-encoded states within a single real algebra, and provides a description of stabilizer dynamics purely in terms of local geometric products. As an equivariance condition, we show that a map intertwines the left Clifford action on the algebra with the induced action on the minimal ideal, actualizing the spinor module as a homogeneous representation of the Clifford unit group.

The practical value of real formulations of quantum theory has been debated \cite{HoffreumonWoods2025Quantum, Renou2021}, with some physicists maintaining that complex numbers are fundamental because the real-number alternatives require "hidden non-local degrees of freedom" or less intuitive mathematical structures to work. Foundationally equivalent to complex formulations, real encodings are often seen as offering no general asymptotic advantage, particularly given the efficiency of stabilizer tableau methods and the maturity of optimized complex linear algebra libraries. While I give reason to believe that this conclusion is not the whole story, the present work does not advocate the real formalism primarily for bit-level compression which is already near minimal \cite{Aaronson2004ImprovedStabilizer, gidney2021stim}, rather, the advantage is structural. Clifford algebraic actions preserve multivector grading and respect the geometric product, enabling invariants to be extracted as grade projections, often as scalar components, without recourse to full matrix evaluation. In particular, quantities such as the trace reduce to scalar projections whose cost is independent of the Hilbert space dimension. This structural transparency aligns with the geometrical algebra formulations of projective geometry and with recent results on Clifford-equivariant architectures \cite{ruhe2023, Nguyen2024EquivariantQNN}, where grade decomposition yields inequivalent subrepresentations and polynomial equivariant layers. Furthermore, due to its inherent relation with stabilizer formalism, this framework suggests a more transparent reinterpretation of the Gottesman-Knill theorem \cite{Gottesman1998Heisenberg} and \textit{Clifford hierarchy} defined by Gottesman and Chuang (1999) \cite{Gottesman1999Teleportation}, in which the Clifford group inherits a direct geometric meaning rather than appearing solely as an abstract normalizer of the Pauli group.

Quantum simulation for universal gate sets has remained constrained by the usual bottlenecks, including exponential state dimension, operator growth under composition, and the nonlocal propagation of phase and entanglement. Although this Clifford algebra representation is linearly equivalent to the usual matrix formalism in the worst case (a generic multivector is as information-dense as a generic density matrix),  the algebraic language—grades, involutions, idempotents, and the internal 'imaginary' unit $J$—renders important families of states and channels visibly low-complexity. The aim is to exploit such to derive hybrid algorithms capable of simulating restricted yet physically meaningful quantum processes more efficiently than naïve matrix-based treatments would suggest.

\subsection{Clifford Algebra and the Qubit Isomorphism}

A Clifford algebra \( \mathcal{C}\ell_{p,q}(\mathbb{K}) \) is an associative algebra over a field \( \mathbb{K} \) generated by a vector space \( V \) of dimension \( n = p + q \) equipped with a quadratic form \( Q: V \rightarrow \mathbb{K} \). It is defined by the fundamental relation $v^2 = Q(v) \cdot 1,   \forall v \in V$,
where \( 1 \) denotes the multiplicative identity. The quadratic form induces the decomposition of the \emph{geometric product} of two vectors as $uv = u \cdot v + u \wedge v$, the sum of a symmetric inner product and an antisymmetric outer product. With the orthogonal basis $\{1, e_i, e_{ij}, \dots, e_{1\cdots n}\}$, the algebra is $\mathbb{Z}_2$-graded with the grade decomposition,
\[
\mathcal{C}\ell_{p,q}(\mathbb{K}) = \bigoplus_{k=0}^n \mathcal{C}\ell_{p,q}^k(\mathbb{K}),
\]
where grade-\(k\) elements represent oriented \(k\)-dimensional subspaces.

Clifford algebras generalize complex numbers and higher-dimensional division algebras, providing the algebra underlying reflections and rotations via exponentials of bivectors, known as \emph{rotors} $R = e^{-\frac{\theta}{2} e_i e_j}$. The complex field \( \mathbb{C} \), with basis \( \{1, i\} \), is isomorphic to the real Clifford algebra \( \mathcal{C}\ell_{0,1}(\mathbb{R}) \), where the pseudoscalar satisfies \( i^2 = -1 \). For quantum information, the single-qubit operator algebra $M_2(\mathbb{C})$ is the Pauli basis,
\[
\text{End}(\mathbb{C}^2) \cong M_2(\mathbb{C}) = \text{span}\{I, \sigma_x, \sigma_y, \sigma_z\}.
\]

\begin{comment}
As a real algebra, this operator space is eight-dimensional; however, its multiplicative structure is already encoded by the four-dimensional real Clifford algebra $\mathcal{C}\ell_{2,0}(\mathbb{R})$, once an internal complex structure is specified. 
\end{comment}

Hestenes demonstrated that real geometric algebras provide a conceptually unified description of quantum mechanics, assigning geometric meaning to and simplifying equations appearing in electromagnetism and Dirac theory \cite{Hestenes2015}. Here we extend this geometric viewpoint from physical interpretation to computation, developing a real formalism for quantum information suitable for classical simulation on modern hardware. Our focus is the real algebra $\mathcal{C}\ell_{2,0}(\mathbb{R})$, which is isomorphic to $M_2(\mathbb{R})$ and becomes isomorphic to $M_2(\mathbb{C})$ once an internal complex structure is introduced.

\begin{comment}
In particular, under the canonical identification
$
\{1, e_1, e_2, e_{12}\} \quad \longrightarrow \quad \{I, \sigma_z, \sigma_x, i\sigma_y \},
$
the basis elements of $\mathcal{C}\ell_{2,0}(\mathbb{R})$ align directly with the Pauli generators. In the following sections, we show how the bivector $J=e_{12}$ functions as an internal imaginary unit and how the full Pauli algebra emerges from the geometric product, establishing the algebraic foundation for State–Operator Compatibility and a scalable, grade-preserving multi-qubit model.
\end{comment}
\section{The State-Operator Clifford Compatibility}
Idempotents provide a canonical mechanism for decomposing any unital associative algebra. If $P$ satisfies $P^2=P$, then its complement $Q:=1-P$ is automatically an orthogonal idempotent. In quantum mechanics, projection operators are ubiquitous as representations of observables and measurement outcomes, yet their decompositions are typically treated at the level of matrix blocks rather than as immanent features of the operator algebra itself \cite{neumann1955mathematical}. Here, our focal point is the primitive idempotent $P$ and its complement $Q$ whose matrix representation is
\[
\rho(P)=\rho\left(\frac{1+e_1}{2}\right)
=\begin{pmatrix}
1 & 0\\
0 & 0
\end{pmatrix},
\qquad 
\rho(Q)=\rho\left(\frac{1-e_1}{2}\right)
=\begin{pmatrix}
0 & 0\\
0 & 1
\end{pmatrix}, 
\]
which forms the spectral decomposition of $e_1$. Algebraically they induce the \textit{Peirce decomposition} \cite{pierce1907linear,herstein1968noncommutative}
\[
\mathcal{C}\ell_{2,0}
=
P\mathcal{C}\ell_{2,0}P
\;\oplus\;
P\mathcal{C}\ell_{2,0}Q
\;\oplus\;
Q\mathcal{C}\ell_{2,0}P
\;\oplus\;
Q\mathcal{C}\ell_{2,0}Q,
\]
which separates diagonal components $P\mathcal{C}\ell_{2,0}P$ and $Q\mathcal{C}\ell_{2,0}Q$ from off-diagonal bimodules $P\mathcal{C}\ell_{2,0}Q$ and $Q\mathcal{C}\ell_{2,0}P$. Unlike block decompositions arising from central idempotents which yield direct product decompositions of algebras, the Peirce decomposition generally retains nontrivial off-diagonal terms and is an additive, rather than multiplicative, decomposition \cite{lam2001first}. Under $\rho$, these are the four matrix units relative to the computational basis,
\[
P^2 = P = a^{\dagger}a , \quad Q^2 = Q = aa^{\dagger},  \qquad aa^{\dagger} + a^{\dagger}a = P + Q=1, \quad PQ=QP =0,
\]
where the off-diagonal Peirce components are nilpotent generators from the remaining elements $\{e_2,J\}$,
\[
\rho(a^{\dagger})=\rho\left(\frac{e_2+J}{2}\right)
=\begin{pmatrix}
0 & 1\\
0 & 0
\end{pmatrix} 
\qquad 
\rho(a)=\rho\left(\frac{e_2 - J}{2}\right)
=\begin{pmatrix}
0 & 0\\
1 & 0
\end{pmatrix}.
\]
\noindent
$\{P,Q,a,a^\dagger\}$ follow the canonical anti-commutation relations
$
a^2=(a^\dagger)^2=0,
\{a,a^\dagger\}=1,
$
and span $Q\mathcal{C}\ell_{2,0}P$ and $P\mathcal{C}\ell_{2,0}Q$, thus forming a complete set of matrix units such that
$
\mathcal{C}\ell_{2,0}(\mathbb{R})
=
\mathrm{span}\{P,Q,a,a^\dagger\}
\cong M_2(\mathbb R).
$

The Peirce decomposition of the Clifford algebra reveals a finite Fock-type decomposition in which diagonal idempotents (occupation projectors) encode measurement sectors, nilpotent ladder operators generate transitions between sectors, and the bivector supplies the intrinsic generator of complex phase. Closely related projector-based decompositions appear throughout quantum theory (e.g., $P$–$Q$ projection methods in many-body physics \cite{feshbach1958unified}), though typically without direct identification as such. This decomposition separates sector-preserving components ($P\mathcal{C}\ell_{2,0}P$, $Q\mathcal{C}\ell_{2,0}Q$) from sector-transition operators ($P\mathcal{C}\ell_{2,0}Q$, $Q\mathcal{C}\ell_{2,0}P$), so that quantum evolution is expressed as algebraic transitions within $\mathcal{C}\ell_{2,0}$. In this paradigm, the qubit is realized as a minimal left ideal, eliminating the need for an external Hilbert space and unifying states and operators within a single framework.

\subsection{Basis States and Operators}
In $\mathcal{C}\ell_{2,0}(\mathbb{R})$ the canonical blade basis is $\{1,e_1,e_2,e_{12}\}$ with $e_{12}:=e_1e_2$, $e_1^2=e_2^2=1$, and $e_1e_2=-e_2e_1$. Write $J:=e_{12}$ so $J^2=-1$. The unit blades generate the finite group $G=\{\pm1,\pm e_1,\pm e_2,\pm e_{12}\}$ under the geometric product. Fix the primitive idempotent $P:=\tfrac12(1+e_1)$ and form the minimal left ideal $\mathcal{S}:=\mathcal{C}\ell_{2,0}P$, which is the real spinor space. To recover the complex two-dimensional state space within the real Clifford algebra, we take the $J$-closure
$
\overline{\mathcal{S}}:=\mathcal{S}\oplus \mathcal{S}J,
$
which is stable under right $J$. Declare the complex structure by $\psi\,i:=\psi\,J$, so $(\overline{\mathcal{S}},\cdot J)\cong\mathbb{C}^2$. Set
\[
\mathbf V_1:=\overline{\mathcal{S}},\qquad
\mathbf P_1:=\mathrm{End}(\overline{\mathcal{S}}),
\]
and define the dual map $\mathcal D=(\vartheta,\rho)$ by
\[
\vartheta:\mathcal{C}\ell_{2,0}\to \mathbf V_1,\quad \vartheta(g):=gP,\qquad
\rho:\mathcal{C}\ell_{2,0}\to \mathbf P_1,\quad \rho(g)(\psi):=g\psi.
\]

\begin{theorem}[State–Operator Clifford Compatibility]\label{thm:compat}
With the above definitions, $\rho$ is an algebra representation and, for all $g,h\in\mathcal{C}\ell_{2,0}$, 
\[
%\boxed{\ 
\rho(g)\,\vartheta(h)=\vartheta(gh).
%\ }.
\]
\end{theorem}

% Define custom styles for the diagram
\tikzset{
    myspace/.style={rectangle, draw, minimum width=3.5cm, minimum height=1cm, align=center},
    myarrow/.style={-{Stealth[scale=1.2]}, very thick}
}

% Define custom styles for the diagram
\tikzset{
    myspace/.style={rectangle, draw, minimum width=3.5cm, minimum height=1cm, align=center},
    myarrow/.style={-{Stealth[scale=1.2]}, very thick},
    labelstyle/.style={text width=4.5cm, align=center} % Style for multi-line labels
}

% Define custom styles for the diagram
\tikzset{
    myspace/.style={rectangle, draw, minimum width=3.5cm, minimum height=1cm, align=center},
    myarrow/.style={-{Stealth[scale=1.2]}, very thick},
    labelstyle/.style={text width=4.5cm, align=center} % Style for multi-line labels
}

\begin{figure}[h!]
    \centering
    \begin{tikzpicture}[x=1.2cm, y=1.2cm] % Reduced scale slightly

        % Define horizontal separation (Hsep) and label offset (LOffset)
        \def\Hsep{5} 
        \def\LOffset{1.25} % Increased offset  to prevent overlap

        % 1. Nodes (The Spaces)
        \node[myspace] (Cl_top) at (0, 0) {
            \textbf{Clifford Space}\\ 
            $\mathcal{C}\ell_{2,0}(\mathbb{R})$
        };
        \node[myspace] (Cl_bottom) at (\Hsep, 0) {
            \textbf{Clifford Space}\\ 
            $\mathcal{C}\ell_{2,0}(\mathbb{R})$
        };
        \node[myspace, fill=gray!10] (S_top) at (0, -3.5) {
            \textbf{Quantum State Module}\\ 
            Minimal Left Ideal $\mathcal{S}$
        };
        \node[myspace, fill=gray!10] (S_bottom) at (\Hsep, -3.5) {
            \textbf{Quantum State Module}\\ 
            Minimal Left Ideal $\mathcal{S}$
        };

        % 2. Arrows (The Transformations)

        % Top Arrow: Algebraic Shortcut (gh)
        \draw[myarrow] (Cl_top) -- (Cl_bottom);
        \node[labelstyle] at (2.5, 0.75) {Geometric Product};
        \node at (2.5, -0.45) {$\times g$};
        \node at (2.5, -0.75) {$h \longrightarrow gh$}; 
        
        % Bottom Arrow: Quantum Evolution (Matrix Action)
        \draw[myarrow, blue] (S_top) -- (S_bottom);
        \node[labelstyle, blue] at (2.5, -2.75) {Unitary Evolution};
        \node[blue] at (2.5, -3.95) {$\rho(g)\psi$};

        % Left Arrows: State Mapping (\vartheta) - COMBINED AND OFFSET
        \draw[myarrow, red] (Cl_top) -- (S_top);
        \node[red, align=center, text width=2.5cm] at (-\LOffset, -1.75) {
            State Map $\vartheta$ \\
            $\vartheta(h) = hP$
        };

        % Right Arrows: Final State Mapping (\vartheta) - COMBINED AND OFFSET
        \draw[myarrow, red] (Cl_bottom) -- (S_bottom);
        \node[red, align=center, text width=2.5cm] at (\Hsep + \LOffset, -1.75) {
            State Map $\vartheta$ \\
            $\vartheta(gh) = (gh)P$
        };

        % 3. The Identity (The Proof)
        % \node[anchor=north] at (2.5, 1.5) {
            % \textbf{State–Operator Clifford Compatibility Theorem (\ref{thm:compat})}
        % };
        % \node[anchor=north, align=center] at (2.5, 1.2) {
            % The diagram commutes, proving the identity:
        % };
        
    \end{tikzpicture}
    \caption{Commutative diagram for the \textbf{State--Operator Clifford Compatibility}. Evolving a state via $\rho(g)$ (bottom path) is equivalent to first composing Clifford elements in $\mathcal{C}\ell_{2,0}$ and then mapping to the state space via $\vartheta$ (top path).}
    \label{fig:compatibility_diagram}
\end{figure}

\begin{proof}
\noindent\textbf{(1) States (spinors).}
By definition $\vartheta(h)=hP\in\mathcal{S}\subset\overline{\mathcal{S}}$. 
Since $J^2=-1$, the rule $(\alpha+i\beta)\cdot\psi:=\alpha\,\psi+\beta\,(\psi J)$ endows $\mathbf V_1$ with a $\mathbb C$–vector space. Associativity gives $(g\psi)J=g(\psi J)$, so each $\rho(g)$ commutes with right multiplication by $J$ and is therefore $\mathbb C$–linear on $\mathbf V_1$.
With the ordered spinor basis
\[
|0\rangle:=P,\qquad |1\rangle:=e_2P,
\]
every $\psi\in \mathbf V_1$ has a unique expansion
$
\psi=(\alpha+\beta J)\,|0\rangle+(\gamma+\delta J)\,|1\rangle
$
with $\alpha,\beta,\gamma,\delta\in\mathbb{R}.$
Define the complex-linear qubit-spinor map
\[
\Upsilon:\mathbf V_1\longrightarrow\mathbb{C}^2,\qquad 
\Upsilon\!\big((\alpha+\beta J)P+(\gamma+\delta J)(e_2P)\big)
=\begin{psmallmatrix}\alpha+i\beta\\[2pt]\gamma+i\delta\end{psmallmatrix}.
\]
Then $\Upsilon(P)=(1,0)^{\mathsf T}$ and $\Upsilon(e_2P)=(0,1)^{\mathsf T}$, so $\Upsilon$ is a complex-linear isomorphism.
Define $\pi:\mathrm{End}_{\mathbb{R}}(\mathbf V_1)\to\mathbb{C}^2$ by 
$
\pi(T):=\Upsilon\!\big(T|0\rangle\big).
$
Consequently,
$
\Upsilon\big(\vartheta(g)\big)=\Upsilon(gP)=\pi\!\big(\rho(g)\big),
$
i.e.\ $\Upsilon(\vartheta(g))$ is the \emph{first column} of the matrix of $\rho(g)$ in the basis $\{P,e_2P\}$.

\medskip
\noindent\textbf{(2) Operators (Pauli).}
Define $\rho(g):\mathbf V_1\to\mathbf V_1$ by left multiplication, so $\rho(g)\in \mathbf P_1=\mathrm{End}_{\mathbb{R}}(\mathbf V_1)$ and its matrix is $[\rho(g)]=\mathrm{Ad}_\Upsilon(\rho(g))$.
Associativity in $\mathcal{C}\ell_{2,0}$ yields $\rho(gh)=\rho(g)\rho(h)$ and $\rho(1)=\mathrm{Id}$, so $\rho$ is an algebra representation.
In the spinor basis $\{|0\rangle:=P,\ |1\rangle:=e_2P\}$, the matrix of $\rho(g)$ is determined by the actions
\[
e_1(P)=P,\quad e_1(e_2P)=-(e_2P),\qquad
e_2(P)=(e_2P),\quad e_2(e_2P)=P,
\]
hence
\[
\rho(e_1)=\begin{pmatrix}1&0\\[2pt]0&-1\end{pmatrix}=\sigma_z,\qquad
\rho(e_2)=\begin{pmatrix}0&1\\[2pt]1&0\end{pmatrix}=\sigma_x,\qquad
\rho(J)=\rho(e_1e_2)=\begin{pmatrix}0&1\\[2pt]-1&0\end{pmatrix}=i\sigma_y.
\]
Consequently $\rho(\mathcal{C}\ell_{2,0})\cong M_2(\mathbb{R})\subset M_2(\mathbb{C})$, and equipping $\mathbf V_1$ with the complex structure induced by right multiplication by $J$ identifies $\mathbf P_1\cong M_2(\mathbb{C})$.

\noindent

\medskip
\noindent\textbf{(3) Compatibility.}
Let $\Phi:\mathcal{C}\ell_{2,0}\to M_2(\mathbb{R})$ denote the standard real–algebra isomorphism (so the matrix of $\rho(g)$ in the spinor basis equals $\Phi(g)$).
For all $g,h\in\mathcal{C}\ell_{2,0}$ and $\psi\in\mathbf V_1$,
\[
\rho(g)\,\vartheta(h)=\vartheta(gh), \qquad
\Upsilon\big(\rho(g)\psi\big)=\Phi(g)\,\Upsilon(\psi).
\]
The first identity follows directly from associativity, $\rho(g)\vartheta(h)=g(hP)=(gh)P=\vartheta(gh)$. 
The second holds because $\rho(g)$ and $\Phi(g)$ represent the same left-multiplication action in the spinor basis, and $\Upsilon$ records the corresponding complex coordinates.

\begin{remark}\label{rem:P-indep}
If $P'=gPg^{-1}$ is any primitive idempotent, then the following $\mathbf V_1',\vartheta',\rho'$ are unitarily equivalent to $(\mathbf V_1,\vartheta,\rho)$ via $\psi\mapsto g\psi$, so the general theory is independent of the particular choice of $P$.
\end{remark}
\end{proof}

\begin{comment}
\begin{corollary}[Faithfulness and irreducibility]\label{cor:faithful}
The representation $\rho:\mathcal C\ell_{2,0}\to \mathrm{End}_{\mathbb R}(\mathbf V_1)$ is faithful and irreducible.
\end{corollary}
If $\rho(g)=0$, then $gP=0$ and $g(e_2P)=0$, so the matrix of $\rho(g)$ in the spinor basis is zero; since $\{P,e_2P\}$ generates the spinor module, this implies $g=0$. For irreducibility, any nonzero complex submodule contains a nonzero vector $\psi$, and acting with suitable Clifford elements yields a nonzero multiple of $|0\rangle$; applying $e_2$ then yields $|1\rangle$, so the submodule equals $\mathbf V_1$.
\end{comment}

The pair $\mathcal{D} = (\vartheta,\rho)$ is a natural equivariant interpretation. 
The algebra $\mathcal{C}\ell_{2,0}$ acts on itself and on the minimal left ideal via left multiplication through $\rho$, while the map $\vartheta$ intertwines these actions, so that multiplication in the algebra descends compatibly to evolution in the spinor module. In this sense, the spinor space is realized as an equivariant image of the algebra. States and operators arise from the same algebra, distinguished only by their roles under the equivariant pair $(\vartheta,\rho)$. 

The same generators play distinct but unified roles-- they create the computational basis elements ($P,e_2P$) via right multiplication by the idempotent $P$ (through $\vartheta$), and act as Pauli operators ($\sigma_z,\sigma_x$) via left multiplication (through $\rho$). The identification $e_1 \leftrightarrow \sigma_z$ is determined by the choice of primitive idempotent $P$ as the physical vacuum projector onto the $+1$ eigenspace of $e_1$. The remaining generators are then fixed by the Clifford relations, at one with the $\mathbb{Z}_2$-grading
\[
\mathcal{C}\ell_{2,0}(\mathbb{R})
=\underbrace{\mathrm{span}\{1,J\}}_{\text{even}}
\oplus
\underbrace{\mathrm{span}\{e_1,e_2\}}_{\text{odd}},
\]
where the even subalgebra $\mathrm{span}\{1,J\}$ governs scalar phase and rotors, while the odd subspace $\mathrm{span}\{e_1,e_2\}$ generates the Pauli axes under $\rho$.

\subsection{Generalization to Multi-Qubit Systems} 
Being $\mathbb Z_2$–graded, the Clifford algebra supports both the ordinary (Kronecker) tensor product of associative algebras and the graded (Koszul) tensor product, in which homogeneous elements super–commute according to
$
(x \widehat{\otimes} y)(x' \widehat{\otimes} y')
=
(-1)^{|y||x'|}(xx') \widehat{\otimes} (yy').
$
Let the $2n$–dimensional quadratic space decompose as an orthogonal direct sum
$
V = V_1 \oplus \cdots \oplus V_n, \ \dim V_i = 2,
$
with each $V_i$ equipped with the induced quadratic form. Then there is a canonical isomorphism of $\mathbb Z_2$–graded algebras
$
\mathcal C\ell(V)
\cong
\mathcal C\ell(V_1)\widehat{\otimes}\cdots\widehat{\otimes}\mathcal C\ell(V_n),
$
and, identifying each $\mathcal C\ell(V_i) \cong \mathcal C\ell_{2,0}(\mathbb R)$,
$
\mathcal C\ell_{2n,0}(\mathbb R)
\cong
\mathcal C\ell_{2,0}(\mathbb R)^{\widehat{\otimes} n}.
$
The distinction between the graded and ordinary tensor products lies in the multiplication rule across factors: the former introduces the Koszul sign so that odd elements anticommute across distinct factors, while the latter preserves locality. The compatibility principle depends only on associativity of the geometric product and therefore remains monoidally invariant under either choice. This flexibility allows the same local algebra to support both qubit-style tensor structure and, under graded composition, fermionic modes, pointing toward occupation-based (Fock-type) constructions.

To model the local tensor-product structure of qubit systems, we use the Kronecker tensor algebra $\mathcal{C}\ell_{2,0}(\mathbb{R})^{\otimes n}$, which preserves locality of generators. The $n$-qubit computational basis vacuum $|0\cdots 0\rangle$ is given by the tensor product of primitive idempotents
\[
P_n:=\bigotimes_{i=1}^n P^{(i)} = \bigotimes_{i=1}^n \tfrac12(1+e_1^{(i)}),
\qquad 
\mathcal S_n:=\mathcal{C}\ell_{2,0}(\mathbb{R})^{\otimes n} P_n,
\]
where $e_1^{(i)}$ acts locally on the $i$-th qubit. The resulting minimal left ideal $\mathcal S_n$ generalizes the single-qubit spinor space, and the full complex space required for the $2^n$–dimensional Hilbert space $\mathbb{C}^{2^n}$ is obtained by $J$-closure,
\[
\mathbf V_n:=\overline{\mathcal S}_n:=\mathcal S_n\oplus \mathcal S_n J_{\mathrm{glob}}, 
\qquad 
\mathbf P_n:=\mathrm{End}(\overline{\mathcal S}_n),
\]
where $J_{\mathrm{glob}}:=J^{(1)}\otimes I^{\otimes(n-1)}$ is a fixed bivector generator with $J_{\mathrm{glob}}^2=-1$; any other single-site choice \(J^{(i)} = e_{12}^{(i)}\) results in an isomorphic right action after permuting tensor factors. Define the $n$-qubit dual maps
\[
\vartheta_n(G):=G P_n,
\qquad 
\rho_n:=\rho^{\otimes n}:\ \mathcal{C}\ell_{2,0}(\mathbb{R})^{\otimes n}\to \mathbf P_n.
\]
\begin{corollary}[Scaled State--Operator Compatibility]
For all $G,H\in \mathcal{C}\ell_{2,0}(\mathbb{R})^{\otimes n}$,
$
\rho_n(G)\,\vartheta_n(H)=\vartheta_n(GH).
$
In particular, for local generators,
\[
\rho(e_1^{(i)})=I^{\otimes(i-1)}\!\otimes \sigma_z \otimes I^{\otimes(n-i)},\quad
\rho(e_2^{(i)})=I^{\otimes(i-1)}\!\otimes \sigma_x \otimes I^{\otimes(n-i)},\quad
\rho(J^{(i)})=I^{\otimes(i-1)}\!\otimes i\sigma_y \otimes I^{\otimes(n-i)}.
\]
\end{corollary}

The multi-qubit Clifford algebra thus provides a direct algebraic realization of Pauli operators. Under the representation $\rho$, the generators $\{e_1^{(i)},e_2^{(i)},J^{(i)}\}$ act as the single-qubit Pauli operators $\{\sigma_z,\sigma_x,i\sigma_y\}$, and tensor products of these elements generate the full set of $n$-qubit Pauli strings. The $2^{2n}$ basis blades of $\mathcal{C}\ell_{2n,0}(\mathbb{R})$ are in bijection with the $2^{2n}$ Pauli operators, providing an algebraic labeling of the operator basis and spanning the associated Lie algebra of Pauli operators, though we defer discussion of exponentiation and unitary generation. This identification allows operators to be read directly from the graded algebra and motivates the grade--weight correspondence developed in the next section.

\subsubsection{Grade--Weight Correspondence}

To maintain algebraic precision, we distinguish between two related measures of operator complexity. The \emph{Pauli weight} $w(A)$ of a Pauli string $A=a_1\otimes\cdots\otimes a_n$ is the number of non-identity factors, a standard measure of support in quantum information theory. The \emph{vector weight} $w_v(A)$ is the total number of grade-1 generators ($e_1$ or $e_2$) appearing in the Clifford algebraic representation of $A$, and coincides with the total algebraic grade of the multivector. At the single-qubit level, the operators $\sigma_z$ and $\sigma_x$ are grade-1 elements, while $\sigma_y$ is the bivector $J=e_1e_2$ and has grade $2$. The two notions are related by
$
w_v(A) = w(A) + n_Y,
$
where $n_Y$ denotes the number of $J$ factors. The algebraic grade is additive across tensor factors. For $A=a_1\otimes\cdots\otimes a_n \in \mathcal{A}_n=\mathcal{C}\ell_{2,0}(\mathbb{R})^{\otimes n}$ ,
\[
\mathrm{grade}(A) = \sum_{i=1}^n \mathrm{grade}(a_i) = w_v(A).
\]
For example, $e_1^{(1)} \otimes e_{12}^{(2)}$ has total grade $3$, reflecting the sum of contributions from each tensor factor. For homogeneous (blade) tensor products, this composition implies that the geometric product of Pauli strings factorizes locally,
$
BA = (b_1a_1)\otimes\cdots\otimes(b_na_n),
$
so that both multiplication and grade evaluation are $n$ independent local operations. Unlike Pauli weight, which captures only support, vector weight encodes the full geometric content of an operator, distinguishing elements that act on the same qubits but differ in algebraic structure (e.g., vector vs bivector contributions), refining support into geometry.

\subsection{Non-Uniqueness of Operator Representatives}

The preceding identifies quantum states with elements of $\mathcal S_n = \mathcal A P_n$, where each algebra element $A \in \mathcal A$ determines a state via the projection $\vartheta_n(A) = A P_n$. However, this representation is not unique, and distinct elements may correspond to the same state in $\mathcal S_n$. In the Hilbert space formalism, states are represented as vectors modulo global phase, with no record of the generating operator from the ground state. We instead see states as elements of the orbit of $P_n$ under the left action of the algebra (or its unit group), with non-uniqueness appearing as freedom in the choice of lift. Non-uniqueness arises because $\vartheta_n$ is not injective; its kernel consists of all elements annihilated upon right multiplication by $P_n$. So, states are best regarded as equivalence classes $[A] = A + \mathcal A Q_n$, where the quotient separates physically meaningful information from algebraic degrees of freedom that act trivially on the state space.

\begin{definition}(Left annihilator).
The left annihilator of $P_n$ is
$
\mathrm{Ann}_L(P_n)
:=
\{B\in \mathcal{A}:\; B P_n = 0\}.
$
\end{definition}
\noindent
Elements of $\mathrm{Ann}_L(P_n)$ are those that vanish under the projection $A \mapsto A P_n$ onto the left ideal $\mathcal A P_n$.

\begin{proposition}[Quotient description of the spinor space]
\label{prop:quotient-description}
The state map $\vartheta$ is a surjective $\mathbb R$-linear map with kernel $\ker(\vartheta)=\mathrm{Ann}_L(P_n)$.
Consequently,
\begin{equation}
\label{eq:ideal-quotient}
\mathcal{S}_n \cong \mathcal{A} / \mathrm{Ann}_L(P_n) = \mathcal{A} /\mathcal{A}Q_n.
\end{equation}
\end{proposition}

\begin{proof}
Surjectivity is immediate from the definition of $\mathcal S_n $.
Moreover $\vartheta(A)=0$ iff $A P_n=0$, i.e.\ iff $A\in\mathrm{Ann}_L(P_n)$.
The isomorphism $\mathcal S_n \cong\mathcal{A} / \mathrm{Ann}_L(P_n)$ follows from the first isomorphism theorem for vector spaces.
Let $Q_n := 1 - P_n$ denote the complementary idempotent. We now show 
$
\mathrm{Ann}_L(P_n) = \mathcal{A} Q_n.
$
Since $P_n$ is idempotent,
$
(1-P_n)P_n = P_n - P_n^2 = 0.
$
Thus every element of $\mathcal A(1-P_n)$ annihilates $P_n$ on the left,
$
A(1-P_n)P_n = A\cdot 0 = 0.
$
Hence $\mathcal A(1-P_n) \subseteq \mathrm{Ann}_L(P_n)$. Conversely, suppose $B P_n = 0$.  Using $1 = P_n + (1-P_n)$,
\[
B = B(P_n + (1-P_n)) = BP_n + B(1-P_n) = B(1-P_n) = BQ_n
\]
Hence $B \in \mathcal{A}Q_n$.
\end{proof}

\begin{remark}
The subspace $\mathcal A Q_n = \mathrm{Ann}_L(P_n)$ is a left ideal complementary to $\mathcal A P_n$, and the quotient $\mathcal A / \mathcal A Q_n$ carries a left $\mathcal A$–module with $A \cdot [B] = [AB]$.
\end{remark}

\begin{corollary}[Algebra decomposition]
\label{cor:direct-sum}
The algebra splits as a direct sum
$
\mathcal A = \mathcal A P_n \;\oplus\; \mathcal A Q_n.
$
\end{corollary}

\begin{proof}
Every $A\in \mathcal A$ decomposes uniquely as
$
A = A P_n + A(1-P_n),
$
with $A P_n \in \mathcal A P_n$ and $A(1-P_n) \in \mathcal A(1-P_n)$.
The two subspaces intersect trivially because if
$
A P_n = B(1-P_n),
$
then multiplying on the right by $P_n$ gives $A P_n = 0$.
\end{proof}

\begin{remark}[Gauge freedom]
\label{rem:gauge}
Two representatives $A$ and $A'$ describe the same state iff
$
A' = A + B,\
B \in \mathcal A(1-P_n).
$
Thus each physical state is an affine coset
$
A + \mathcal A(1-P_n).
$ For any site $i$, $Q^{(i)} P_n = 0$ (since $Q^{(i)} P^{(i)} = 0$), and $(1 + Q^{(i)}) P_n = P_n$. Local annihilators are instances of the redundancy, even single-site null directions act trivially on the global state.
\end{remark}

\begin{lemma}[Single-qubit annihilator]
\label{lem:single-qubit-ann}
$
\mathrm{Ann}_L(P) = \mathrm{span}\{Q,\; e_2 Q\}.
$
\end{lemma}

\begin{proof}
We have $Q P=0$ and $(e_2 Q)P = e_2(QP)=0$, so $\mathrm{span}\{Q,e_2Q\}\subseteq\mathrm{Ann}_L(P)$. Since $\dim_{\mathbb R}\mathcal{C}\ell_{2,0}=4$ and $\dim_{\mathbb R}(\mathcal{C}\ell_{2,0}P)=2$, it follows that $\dim \mathrm{Ann}_L(P)=2$. The elements are linearly independent, and form a basis.

\end{proof}
\noindent
The relation $1 \sim e_1$ is not accidental. More generally, Lemma~\ref{lem:single-qubit-ann} shows that for any real parameters $\alpha,\beta$,
\[
1 \sim 1+\alpha Q+\beta e_2 Q,
\qquad
e_1 \sim e_1+\alpha Q+\beta e_2 Q,
\]
since $(\alpha Q+\beta e_2 Q)P = 0$. Thus there is a two-dimensional affine family of distinct operator representatives corresponding to the same physical state in the one-qubit case. In the quotient, the one-qubit ideal basis elements are equivalence classes
\[
|0\rangle = [1] = [e_1], 
\qquad 
|1\rangle = [e_2],
\]
where brackets denote equivalence under $\sim$. This equivalence is useful, as in fact the identification $[1]=[e_1]$ permits multiple representations of the same state, enabling simplifications in computation by selecting representatives that minimize support or reduce multiplication cost.

\subsubsection{Group action and orbit.}

Let $\mathcal G \subset \mathcal A^\times$ be a subgroup acting on the spinor space by left multiplication,
$
G \cdot \psi := G \psi.
$
In particular, $\mathcal G$ acts on the vacuum spinor $P_n$, and its orbit is
$
\mathcal O_{P_n} := \{ G P_n : G \in \mathcal G \} \subseteq \mathcal S_n.
$
For appropriate choices of $\mathcal G$, this orbit generates the relevant class of normalized spinors.

The \emph{isotropy subgroup} of $P_n$ in $\mathcal G$ is
\[
\mathrm{Iso}_{\mathcal G}(P_n)
\;:=\;
\{\, G \in \mathcal G \;:\; G P_n = P_n \,\}.
\]
Elements of $\mathrm{Iso}_{\mathcal G}(P_n)$ act trivially on the vacuum spinor and represent gauge transformations relative to $P_n$. The orbit $\mathcal O_{P_n}$ is identified with the homogeneous space $\mathcal G / \mathrm{Iso}_{\mathcal G}(P_n)$. Define an equivalence relation on $\mathcal G$ by
$
G \sim H
\Longleftrightarrow 
G^{-1} H \in \mathrm{Iso}_{\mathcal G}(P_n),
$
so the equivalence classes are the left cosets,
$
[G] = G\,\mathrm{Iso}_{\mathcal G}(P_n)\in \mathcal G / \mathrm{Iso}_{\mathcal G}(P_n).
$
The map
\[
\varphi : \mathcal G / \mathrm{Iso}_{\mathcal G}(P_n) \longrightarrow \mathcal O_{P_n},
\qquad
\varphi\bigl(R\,\mathrm{Iso}_{\mathcal G}(P_n)\bigr) = R P_n,
\]
is well-defined (independent of the choice of representative) and bijective, thus
$
\mathcal O_{P_n} \cong \mathcal G / \mathrm{Iso}_{\mathcal G}(P_n)
$
as sets (and, in suitable settings, as manifolds).

\begin{lemma}
For $A,B \in \mathcal G$,
\[
A P_n = B P_n
\quad\Longleftrightarrow\quad
B^{-1} A \in \mathrm{Iso}_{\mathcal G}(P_n).
\]
\end{lemma}

Thus the fibres of $\vartheta$ are precisely the left cosets of the isotropy subgroup,
$
\vartheta^{-1}(A P_n) = A\,\mathrm{Iso}_{\mathcal G}(P_n),
$
and $\vartheta$ factors through the quotient,
\[
\mathcal G \;\twoheadrightarrow\; \mathcal G / \mathrm{Iso}_{\mathcal G}(P_n) \;\xrightarrow{\;\varphi\;}\; \mathcal O_{P_n} \subseteq \mathcal S_n.
\]
Since $P_n$ is non-invertible, the map $\vartheta$ is not invertible, so recovering an operator from a spinor requires a choice of representative.

\begin{definition}
A \emph{lift} of a spinor $\psi \in \mathcal O_{P_n}$ is an element $G \in \mathcal G$ such that
$
G P_n = \psi.
$
\end{definition}

Since $\psi$ corresponds to a coset $G\,\mathrm{Iso}_{\mathcal G}(P_n)$, lifts are not unique. If $G P_n = \psi$ and $H \in \mathrm{Iso}_{\mathcal G}(P_n)$, then $(GH) P_n = \psi$. A \emph{canonical lift} is a section
\[
\sigma : \mathcal O_{P_n} \longrightarrow \mathcal G,
\qquad
\sigma(\psi) P_n = \psi,
\]
chosen by a prescribed gauge (e.g.\ minimal grade or fixed phase). Composing $\sigma$ with $\vartheta$ yields the identity on the orbit,
$
\vartheta\bigl(\sigma(\psi)\bigr) = \psi,
\; 
\forall\,\psi \in \mathcal O_{P_n}.
$ In this language, lifting a state amounts to selecting a representative of its coset in $\mathcal G / \mathrm{Iso}_{\mathcal G}(P_n)$.

\medskip
In the single-qubit case, any normalized spinor can be written in the gauge-fixed representation
$\psi = \alpha P + \beta e_2 P = (\alpha + \beta e_2) P,
\; \alpha,\beta \in \mathbb R,\; \alpha^2 + \beta^2 = 1. $
For $n>1$ such linear parameterizations no longer suffice, a general lift $A \in \mathcal{C}\ell_{2,0}^{\otimes n}$ may contain higher-grade components and terms spanning multiple tensor factors. Factorized lifts,
\[
G = \bigotimes_i (\alpha_i + \beta_i e_2^{(i)}),
\]
are separable states, while entangled states require non-factorizable multivectors.

For geometric operations, it is often preferable to use a \emph{rotor-aligned lift}, a unit rotor $R_\theta \in \mathcal G$ such that $R_\theta P = \psi$.
For separable states, one may take
\[
R = \bigotimes_i (\alpha_i - \beta_i J^{(i)}),
\]
with $\alpha_i = \cos(\theta_i/2)$ and $\beta_i = \sin(\theta_i/2)$, which produces the same spinor as $G$ under projection but is adapted to Bloch-sphere geometry and conjugation. Canonical lifts provide minimal coordinates in the ideal, while rotor-aligned lifts generate observable geometry.

The isotropy subgroup $\mathrm{Iso}_{\mathcal G}(P_n)$ is the stabilizer of the vacuum, and for $\psi = AP_n$,
$
\mathrm{Iso}_{\mathcal G}(\psi) = A\,\mathrm{Iso}_{\mathcal G}(P_n)\,A^{-1},
$
so all stabilizers are obtained by conjugation of the subgroup. What follows is the characterization
\[
\mathrm{Iso}_{\mathcal G}(P_n)
=
\{\, G \in \mathcal G : (G-1)P_n = 0 \,\}
=
\mathcal G \cap \bigl(1 + \mathrm{Ann}_L(P_n)\bigr).
\]
The freedom in choosing a lift $A$ of a state $\psi = AP_n$ is  right multiplication by elements whose deviation from the identity lies in $\mathrm{Ann}_L(P_n)$. This condition fixes the first column of $G$ while leaving the remaining blocks unconstrained up to invertibility, so the action on the ideal depends only on a restricted subset of components.

\begin{comment}
States therefore correspond to cosets in $\mathcal G / \mathrm{Iso}_{\mathcal G}(P_n)$, with representatives retained explicitly. Under evolution by $G \in \mathcal G$,
\[
A P_n \longmapsto (G A) P_n,
\]
so updates reduce to geometric multiplication in $\mathcal A$, with stabilizers transported by conjugation and redundancy controlled by the annihilator of $P_n$.
\end{comment}

\section{Geometric Representations}
In the matrix orthodoxy of quantum mechanics, the complex number $i$ appears in two different ways. It generates global or relative phase factors $e^{i\phi}$ in state amplitudes, and appears intrinsically in operators such as Hamiltonians and Pauli matrices. Although these roles are encoded within the same complex unitary convention, multiplication of a state by a global phase commutes with all observables and has no physical effect, whereas application of a nontrivial unitary operator generates genuine dynamical evolution. 

Within the SOCC, this distinction becomes clear. Physical operators act by left multiplication in the algebra, whereas complex phase arises from a \emph{right action} generated by the bivector $J := e_{12}$ with $J^2 = -1$ after passing to the $J$-closed module. Right multiplication by the rotor $\cos\phi + J\sin\phi$ corresponds to multiplication by the complex phase $e^{i\phi}$. More thoroughly, the spinor space $\mathcal S = \mathcal{C}\ell_{2,0} P$ is a minimal left ideal that carries a (real) right action by the corner algebra $P\mathcal{C}\ell_{2,0}P \cong \mathbb{R}$. In the multi-qubit setting, although each tensor factor contributes a local bivector generator $J^{(i)} = e_{12}^{(i)}$, the underlying right action remains one-dimensional over $\mathbb{R}$. The $J$-closure $\overline{\mathcal S}_n := \mathcal S_n \oplus \mathcal S_n J_{\mathrm{glob}}$ forms an $(\mathcal{A},\;\mathbb{R}[J_{\mathrm{glob}}]/(J_{\mathrm{glob}}^2+1))$-bimodule, with right action generated by $J_{\mathrm{glob}}$, where $\mathbb{R}[J_{\mathrm{glob}}]/(J_{\mathrm{glob}}^2+1)\cong\mathbb{C}$. This separation of left and right actions provides a geometric interpretation of the complex structure underlying qubit states and allows phase information to be tracked independently of operator evolution within the real algebra.

%%%% Change these parameters to change the position of psi, or the size/rotation of the sphere
\def\rotationSphere{-110}
\def\psiLat{45}
\def\psiLon{45}
\newlength{\radiusSphere}
\setlength{\radiusSphere}{2cm}
\begin{figure}[hbt!]
\centering
\begin{blochsphere}[radius=\radiusSphere,opacity=0,rotation=\rotationSphere]
  \drawBallGrid[style={opacity=.2}]{30}{45}
  % Draw the sphere...
  \drawLongitudeCircle[]{\rotationSphere} % draw the longitude that face us to delimit the sphere
  % ... and the equatorial plane
  \drawLatitudeCircle[style={dashed}]{0}
  % Define the different points on the bloch sphere
  \labelLatLon{ket0}{90}{0};
  \labelLatLon{ket1}{-90}{0};
  \labelLatLon{ketminus}{0}{180};
  \labelLatLon{ketplus}{00}{0};
  \labelLatLon{ketpluspi2}{0}{-90};  % Longitude seems to be defined in the "wrong" direction, hence the minus
  \labelLatLon{ketplus3pi2}{0}{-270};
  \labelLatLon{psi}{\psiLat}{-\psiLon};
  % Draw and label the axis
    \draw[-latex] (0,0) -- (ket0) node[above,inner sep=.5mm] at (ket0) {\(P\)};
    \draw[-latex] (0,0) -- (ket1) node[below,inner sep=.5mm] at (ket1) {\(e_2P\ \)};
  \draw[-latex] (0,0) -- (ketplus) node[below,inner sep=.5mm] at (ketplus) {};
  \draw[-latex] (0,0) -- (ketpluspi2) node[below,inner sep=.5mm] at (ketpluspi2) {};
  % Draw |psi>
  \draw[-latex] (0,0) -- (psi) node[above]{ ${\psi}$};

  % Draw the angles
  \coordinate (origin) at (0,0);
  {
    % Will draw the angle/projection one the equatorial plane
    \setDrawingPlane{0}{0}
    
    % Draw the projection: cos is used to compute the length of the projection
    \draw[current plane,dashed] (0,0) -- (-90+\psiLon:{cos(\psiLat)*\radiusSphere}) coordinate (psiProjectedEquat) -- (psi);
    % Draw the angle
    \pic[current plane, draw,fill=orange!50,fill opacity=.5, text opacity=1,"\footnotesize $\phi$", angle eccentricity=2.2]{angle=ketplus--origin--psiProjectedEquat};
     % Draw \(e_{12}\) rotation arrow in the equatorial plane

  }
  { \setLongitudinalDrawingPlane{\psiLon}
    % Draw the angle
    \pic[current plane, draw,fill=orange!50,fill opacity=.5, text opacity=1,"\footnotesize $\theta$", angle eccentricity=1.5]{angle=psi--origin--ket0};
  }

  { \setLatitudinalDrawingPlane{0}
    
    \fill[current plane, fill=orange!20, opacity=0.5] (0,0) circle (\radiusSphere);
       % Draw two \(e_{12}\) rotation arrows outside the sphere on the equatorial plane using \draw
    \draw[current plane,|->, thick, orange, mark size=1.5pt] (2.7,0) arc[start angle=0, end angle=160, radius=2.7];
        \draw[current plane,<-|, thick, orange, mark size=1.5pt] (-2.7,0) arc[start angle=180, end angle=340, radius=2.7];
  }

  \node at (2.35,0.05){{$J$}};
  \node at (-2.35,0) {{$-J$}};

\end{blochsphere}

\caption{Visualization of the Bloch sphere with Clifford algebra labels. \( P \) is \( |0\rangle \) at the north pole, \( e_2P \) is \( |1\rangle \) at the south pole, and \( J \) represents rotations parallel to the equatorial plane.}
\label{fig:clifford_bloch_sphere}
\end{figure}

Following Ablamowicz--Fauser \cite{AblamowiczFauser2011Transposition}, the distinguished transposition anti-involution reduces to
reversion in Euclidean signature, which also satisfies $\widetilde{A} = A^\dagger$ with respect to the Hilbert space inner product under $\rho$.  The Bloch vector $\mathbf{n}$ has an algebraic representation by Clifford conjugation. Let $e_1$ denote the reference observable. Then, in a fixed phase gauge,
\[
\mathbf{n}(\psi)
=
R_\theta\, e_1\, \widetilde{R_\theta},
\]
where $R_\theta$ is a unit rotor lift representing $\psi$. For
$ R_\theta = \cos(\theta/2) - \sin(\theta/2)\,J, $
one computes
$ R_\theta e_1 \widetilde{R_\theta}
=
\cos\theta\, e_1
+
\sin\theta\, e_2,
$
which is a rotation in the $e_1$--$e_2$ plane. The azimuthal degree of freedom is motion in the transverse plane $\mathrm{span}\{e_2,J\}$ and is encoded at the spinor level through the right $J$-action, giving the transverse resolution
$
e_2 \longmapsto \cos\phi\, e_2 + \sin\phi\, J.
$
Since $\mathrm{span}_{\mathbb{R}}\{e_1,e_2,J\}$ is a three-dimensional real vector space, it is (linearly) isomorphic to $\mathbb{R}^3$, and
the full Bloch vector is
\begin{equation}
\mathbf{n}(\theta,\phi)
=
\cos\theta\, e_1
+
\sin\theta\cos\phi\, e_2
+
\sin\theta\sin\phi\, J,
\label{bloch-vector}
\end{equation}
which is the standard sphere parameterization. 

This degree of freedom is implemented by the right action of the bivector generator $J$, with relative phase given by the \textit{right rotor}
$
R = e^{\phi J}.
$
Accordingly, the general normalized pure qubit state with Bloch angles $(\theta,\phi)$ is
\begin{equation}
\psi(\theta,\phi)
=
\cos\!\left(\frac{\theta}{2}\right) P
\;+\;
\sin\!\left(\frac{\theta}{2}\right)\,(e_2P)\,R.
\label{eq:pure-ideal-bloch}
\end{equation}
Expanding the rotor,
$
(e_2P)e^{\phi J}
=
\cos\phi\,(e_2P)+\sin\phi\,(e_2PJ),
$
which shows $\overline{\mathcal{S}}$ is closed under right multiplication by $\mathrm{span}_{\mathbb R}\{1,J\}$, with coefficients encoding both real amplitudes and internal phase. While the rotor acts uniformly on the spinor, only its relative action between the $P$ and $e_2P$ components is observable, reproducing the usual complex phase. Applying the spinor map $\Upsilon$ (first-column extraction in a fixed matrix representation) gives
$
\Upsilon(\psi(\theta,\phi))
=
\cos(\theta/2)\,|0\rangle
+
e^{i\phi}\sin(\theta/2)\,|1\rangle,
$
up to a global phase.

Recall that in Clifford algebra, reflection in a unit vector $v$ is the sandwich map
$
\psi \longmapsto -\,v \psi v,
$
which reverses the component of $\psi$ parallel to $v$ and leaves the orthogonal complement invariant.  More generally, any unit versor $R$ acts by an adjoint transformation
$
\psi \longmapsto R \psi \tilde R,
$
and restricted to the grade-1 subspace, preserves the quadratic form an defines an orthogonal transformation of the underlying vector space. In two dimensions, the unique unit bivector acts on vectors by the adjoint transformation
\[
\psi \longmapsto J \psi (-J).
\]
A direct computation yields
$
J e_1 (-J) = -e_1, 
J e_2 (-J) = -e_2,
$
hence $J \psi (-J) = -\psi$ for all grade-1 elements $\psi$.  Sandwiching by $J$ produces a half-turn (rotation by $\pi$) within the plane where scalars and the bivector $J$ are fixed, while all vectors are negated.  Equivalently, the adjoint action of the bivector is the composition of two reflections across orthogonal vectors spanning the plane (first across $e_1$, then across $e_2$), which in two dimensions becomes central inversion. This geometric picture explains the phase of the Pauli operators. Since $e_2 e_1 = -J$, we obtain
$
Y(\psi) = -\,J \psi J,
$
so the Pauli $\sigma_y$ gate is a composite reflection, producing a half-turn in the $e_{12}$ plane together with the phase on a spinor.
\[
\begin{aligned}
Y^2(\psi) 
  &= -J(-J\psi J)J 
   = J^2 \psi J^2 
   = \psi,\\[4pt]
Y(P) 
  &= -J P J 
   = -(J P)J 
   = -(-e_2 P)J 
   = e_2 P J =: i|1\rangle,\\[4pt]
Y(e_2 P) 
  &= -J (e_2 P) J 
   = -(J e_2 P)J 
   = -P J  =: -i|0\rangle.
\end{aligned}
\]
so $\sigma_y^2 = I$, and $\sigma_y$ acts as expected on the computational basis.

\subsubsection{Density Operators and Compatibility}

A (pure) quantum state is represented as an ideal element $\psi \in \mathcal{S}_n$, and the associated density operator is the rank-one projector
$
\Pi_\psi := \rho(\psi)\,\widetilde{\rho(\psi)}.
$
Since reversion is an anti-involution, $\widetilde{(AP_n)}=\widetilde{P_n}\,\widetilde{A}=P_n\widetilde{A}$, and
\begin{equation}
\Pi_\psi
= \rho(AP_n)\,\widetilde{\rho(AP_n)}
= \rho(AP_n)\,\rho(P_n\widetilde{A})
= \rho(AP_n\widetilde{A}),
\label{eq:pure-density-conjugation}
\end{equation}
showing that pure-state density operators are representations of conjugates of the vacuum idempotent $P_n$. For a single qubit, if $a$ is a unit lift, we obtain
$
\Pi_{\psi}
=
\frac{1}{2}\bigl(1 + a e_1 \widetilde{a}\bigr)
=
\frac{1}{2}\bigl(1 + \mathbf{n}\bigr),
$
where $\mathbf{n}$ is the Bloch vector \eqref{bloch-vector}. For $n$ qubits, this generalizes to a multivector expansion whose grade-1 components give Bloch vectors and whose higher-grade components encode multi-qubit correlations. Thus the Bloch vector is the first component of a higher-dimensional geometric object, with the density operator as its representation under $\rho$. Mixed states are convex combinations of such projectors,
\begin{equation}
\sum_i p_i\,\Pi_{\psi_i}
= \sum_i p_i \rho(A_i P_n \widetilde{A_i}),
\qquad p_i\ge 0,\ \sum_i p_i=1.
\label{eq:mixed-density}
\end{equation}
Equations \eqref{eq:pure-density-conjugation}--\eqref{eq:mixed-density} unify the two viewpoints, density operators are algebra elements, and pure states arise by idempotent conjugation. If $\psi = X P_n$ and an element $A$ acts to produce $\psi' = \rho(A)\psi$, then the density operator evolves by Clifford conjugation,
\[
\Pi_{\psi'} = \rho(A)\,\Pi_\psi\,\rho\widetilde{(A)}
= \rho\big(A (X P_n \widetilde{X}) \widetilde{A}\big)
= \rho\big((AX) P_n \,\widetilde{(AX)}\big).
\]
Thus Schr\"odinger evolution of the generator $X \mapsto AX$ and Heisenberg evolution of the density operator are two manifestations of the same algebraic operation in $\mathcal{C}\ell_{2,0}^{\otimes n}$. For Clifford circuits, the update $X \mapsto AX$ is computed symbolically via local geometric products (bitwise XOR with phase tracking), consistent with stabilizer tableau updates, with $O(n)$ algebraic operations per gate in this representation. Rather than evolving density operators explicitly, one may evolve only the generator and recover observables or density data at the end via a single conjugation $A P_n \tilde{A}$. In closed systems, this avoids operator propagation and reduces simulation to algebraic updates on the generating multivector under local geometric products.

Reversion does not, in general, coincide with inversion, for a generic multivector $A$, one has $A\tilde{A} \neq 1$. 
However, for unit versors,
$
A\tilde{A} = 1, 
\,
A^{-1} = \tilde{A},
$
and no additional normalization is required. Suppose $A$ is stored as a linear combination of basis blades with $c$ nonzero coefficients. 
Computing the reversion $\tilde{A}$ requires only predetermined sign changes. Since reversion is the anti-involution defined by $\widetilde{ab}=\widetilde{b}\,\widetilde{a}$ and $\widetilde{a_k}=(-1)^{\frac{k(k-1)}{2}} a_k$ for homogeneous elements $a_k$ of grade $k$, the cost of reversion is linear in the number of stored coefficients, $O(c)$. The product $A P_n \tilde{A}$ is then obtained by a single geometric product followed by scalar extraction. For comparison, representing the same operator as a dense $2^n\times 2^n$ matrix incurs exponential storage and multiplication costs, with matrix–matrix products requiring $O(2^{3n})$ operations and matrix–vector applications requiring $O(2^{2n})$ operations. 

The matrix trace coincides with a Clifford trace functional given by the scalar component of the multivector. If $A \in \mathcal{C}\ell_{2,0}^{\otimes n}$, then
$
\mathrm{tr}(\rho(A)) = 2^n \langle A\rangle_0,
$
where $\langle A\rangle_0$ denotes the scalar part of $A$. The Born rule \cite{Born1926Zur, Sakurai2020} is expressed directly in Clifford form; for a state $\psi = X P_n$ and observable $A \in \mathcal{C}\ell_{2,0}^{\otimes n}$,
\[
\langle\psi|\rho(A)|\psi\rangle 
= \mathrm{tr}(|\psi\rangle\langle\psi|\,\rho(A)) 
= \mathrm{tr}(\rho(X P_n \widetilde X\, A)) 
= 2^n \langle X P_n \widetilde X\, A\rangle_0.
\]

\medskip

A further computational consequence follows from this identification. Since $\mathrm{tr}(\rho(A))$ is obtained by extracting the scalar component of $A$, evaluating traces or expectation values amounts to reading a single coefficient in the stored multivector expansion. Once the multivector representation is available, scalar extraction is a constant-time operation independent of the Hilbert space dimension, provided coefficients are stored to allow direct access to the scalar component. Grade and Peirce projections similarly act by selecting predetermined coefficient subsets, so their cost depends only on the size of the stored representation. The overall efficiency depends on the sparsity and compression of the multivector representation; when the number of active coefficients remains small, this approach avoids the exponential scaling associated with dense matrix methods.
\section{Endomorphisms of the $J$–Closure}

The definition of physical operators $\mathbf{P}_n$ as \emph{endomorphisms} of the spinor bimodule rather than as elements of the Clifford algebra is intentional.  A local multivector $a\in\mathcal{C}\ell_{2,0}$ acts on a spinor by left multiplication,
$
L_a(\psi)=a\psi,
$
and these left actions form the quantum Clifford group when $a$ is a versor.  However, the spinor space carries an additional right module generated by the bivector, and because Clifford multiplication is associative, left and right actions commute as operators,
\[
L_aR_b(\psi)=a(\psi b)=(a\psi)b=R_bL_a(\psi)
\quad \forall a,b\in\mathcal{C}\ell_{2,0}.
\]
A general linear operator on $\overline{\mathcal S}$ may involve both left and right multiplication, \ a \emph{biaction}
$
\psi \mapsto \sum_i L_{a_i}R_{b_i}(\psi),
$
so left actions correspond to physical Clifford operations, whereas right actions correspond to internal complex phase or state-projection.  This distinction parallels the operator/scalar separation in Hilbert space, but here it is enforced explicitly-- right-$J$ is never absorbed into the left side of a multivector without changing physical meaning.  As such, not all unitary endomorphisms of $\overline{\mathcal S}$ can be represented by a single left multiplier.

\subsubsection*{A universal gate set}

The term ``Clifford'' is now pervasive in quantum computing, where it typically refers to the group of unitary operators that normalize the Pauli group. While this nomenclature honors William Kingdon Clifford, its usage in quantum information is often presented independently of its geometric origins. Historically, however, the Clifford group was defined within Clifford (geometric) algebra as the subgroup of invertible versors (also known as the Lipschitz group), namely those elements whose adjoint action preserves the grade-1 subspace \cite{LundholmSvensson2009}. We adopt this geometric perspective as primary, viewing the quantum Clifford group as a representation.

\begin{lemma}[Versor realization of Pauli normalizers.] Let $\mathcal V$ denote the grade-1 subspace of $\mathcal{C}\ell_{2,0}^{\otimes n}$ spanned by the generators $\{e_1^{(i)}, e_2^{(i)}\}_{i=1}^n$. The subgroup \begin{equation} \Gamma := \{ g \in \mathcal{C}\ell_{2,0}^{\otimes n}{}^{\times} \mid g\,\mathcal V\,g^{-1} \subseteq \mathcal V \} \end{equation} coincides with the group of invertible versors (the Lipschitz/Pin group). A discrete subgroup of $\Gamma$ acts on the spinor space as, up to phase, the normalizer of the Pauli operators; its induced projective action is isomorphic to the $n$-qubit quantum Clifford group. In particular, the standard generating set $\{H,S,\mathrm{CNOT}\}$ is realized by left Clifford multiplication together with the right-$J$ on $\overline{\mathcal S}_n$. \end{lemma}

\begin{remark}
Since $J = e_1 e_2$, preservation of the grade-1 subspace implies $g J g^{-1} = \pm J$ in the single-qubit case. Thus the full Pauli algebra is preserved projectively,  including operators involving the right-$J$.
\end{remark}

This versor characterization realizes the quantum Clifford group inside $\mathcal{C}\ell_{2,0}^{\otimes n}$ by explicit algebra elements. We now give concrete generators for $\mathcal C_2$, expressing $H$, $S$, and $\mathrm{CNOT}$ in terms of Clifford generators and complementary idempotents.

\begin{equation}
H := \frac{e_1+e_2}{\sqrt2}, \qquad 
\text{CNOT}_{c \rightarrow t} := P^{(c)} + Q^{(c)}e_2^{(t)}, \qquad
S := P + QJ.
\label{c2}
\end{equation}
Multi-qubit operations such as $\mathrm{CNOT}$ introduce states whose lifts do not factorize across tensor components. For example, the Bell state $\tfrac{1}{\sqrt{2}}(|00\rangle + |11\rangle)$ is
\[
\Phi^+ = \mathrm{CNOT}_{c \to t}\, H^{(c)}\, P_2,
\qquad
\sigma(\Phi^+) = \frac{1 + e_2^{(c)} e_2^{(t)}}{\sqrt2},
\]
and entanglement is encoded by cross-site blades spanning multiple tensor factors in the lift.  While $H$ and $\text{CNOT}$ are multivectors as left actions, the $S$-gate is more precisely the map
$
S(\psi)=P\psi+Q\psi J.
$
This operator acts as identity on the $P$--sector and applies a right-$J$ phase to the $Q$--sector.  Its adjoint is
$
\widetilde{S}(\psi)=P\psi-Q\psi J,
$
and the orthogonal idempotent relations give immediately
\[
\begin{aligned}
S\widetilde{S}(\psi)
&= P(P\psi-Q\psi J)+Q(P\psi-Q\psi J)J \\
&= P^2\psi-PQ\psi J+QP\psi J-QQ\psi JJ \\
&= P\psi - Q\psi J^2 \\
&= P\psi + Q\psi \\
&= (P+Q)\psi = \psi,
\end{aligned}
\]
and similarly for $\widetilde{S}S(\psi)$, so $S$ is unitary. Although phase gates like the $S$-gate are not representable as single multivectors, they are nonetheless a norm-preserving automorphism of $\overline{\mathcal S}$.  Moreover,
$
S^2(\psi)=P\psi-Q\psi=(P-Q)\psi=e_1\psi,
$
so $S^2$ becomes the operator $\sigma_z = e_1$. It can be checked that 
$
Se_2S^\dagger = J\psi J$ and $Se_1S^\dagger = e_1,
$
thus $S$ normalizes the Pauli algebra and lies in the quantum Clifford group. Quantum Clifford gates are those biactions whose left action preserves the grade-1 subspace and whose right action preserves the distinguished right-$J$ projectively.

In general, higher-level phase gates maintain this pattern in which we replace $J$ with a right rotor. The introduction of rotors (phase angles) marks a boundary of the quantum Clifford group. Let $R=e^{J\pi/4}=\frac{1}{\sqrt2}(1+J)$ be a unit bivector rotor. $\mathcal{C}_3$ contains the Toffoli and $T$-gate as
\begin{equation}
\text{TOFF}_{c_1c_2\rightarrow t} := P^{(c_1c_2)} + Q^{(c_1c_2)}e_2^{(t)}, \qquad T:=P+QR.
\label{c3}
\end{equation}
From the map $T(\psi)=P\psi+Q\psi\,R $, $T$ is unitary and satisfies $T^2=S$.  
Unlike $S$, its conjugation action enlarges the operator span, i.e.
$
Te_2\widetilde{T}(\psi)=\frac1{\sqrt2}(e_2\psi -J \psi J),
$
marking the transition from Clifford to "non–Clifford" behavior. This biaction preserves the Peirce decomposition of the state space but introduces a phase-shear that mixes Pauli operators under conjugation.

Notably, each level $k\leq 3$ of the hierarchy possesses a canonical biaction. For $\mathcal{C}_1 = \{X, Y, Z\}$, $Y$ appears as the adjoint action of $J$ to implement complex structure. For the $\mathcal{C}_2$ generators \eqref{c2}, the $S$ gate introduces the first idempotent-controlled use of this same complex structure. For $\mathcal{C}_3$ \eqref{c3}, the $T$ gate refines this by substituting
the $J$-action for the dyadic rotor, replacing the quarter-turn rotation in the $Q$-sector by an
eighth-turn rotation. More generally, higher levels of the hierarchy allow
idempotent-controlled right-$J$ rotors with dyadic angles
$
\theta=\frac{\pi}{2^m},
$
whose adjoint actions map the Pauli lattice into progressively larger
normalizers inside the versor group. This characterization was identified by Cui and Gottesman \cite{cui2017diagonal}, and is here given a direct geometric interpretation.

\subsubsection{Sector--Rotor Normal Forms and Idempotent Pruning}

An advantage of the idempotent representation is that Clifford$+T$ gates admit a \emph{sector--rotor} normal form that is stable under composition. For example, a $\mathrm{CNOT}_{1\to2}$ followed by a $T$ gate on the control qubit preserves this two-term form under composition. Since $R^{(1)}$ acts only on site $1$ and $e_2^{(2)}$ acts only on site $2$, both commute with idempotents on the other site.

\begin{align*}
U =
\mathrm{CNOT}_{1\to2}\, T_1
& =
\big(P^{(1)} + Q^{(1)} e_2^{(2)}\big)
\big(P^{(1)} + Q^{(1)} R^{(1)}\big). \\
&=
P^{(1)}P^{(1)}
+ P^{(1)} Q^{(1)} R^{(1)}
+ Q^{(1)} e_2^{(2)} P^{(1)}
+ Q^{(1)} e_2^{(2)} Q^{(1)} R^{(1)}.\\
&=
P^{(1)} + Q^{(1)} e_2^{(2)} R^{(1)}.
\end{align*}

The middle two terms vanish since $Q^{(1)}$ and $P^{(1)}$ both act on site $1$, and for the final term, since $Q^{(1)}$ acts only on site $1$,
$
Q^{(1)} e_2^{(2)} Q^{(1)} R^{(1)}
=
(Q^{(1)})^2 e_2^{(2)} R^{(1)}
=
Q^{(1)} e_2^{(2)} R^{(1)}.
$ This form is stable under further $T_1$ gates. Let
$
T_1' := P^{(1)} + Q^{(1)} R'^{(1)}
$
be another $T$-type gate on qubit $1$.  Then
\begin{align*}
U T_1'
&=
\big(P^{(1)} + Q^{(1)} e_2^{(2)} R^{(1)}\big)
\big(P^{(1)} + Q^{(1)} R'^{(1)}\big) \\
&=
P^{(1)}P^{(1)}
+ P^{(1)} Q^{(1)} R'^{(1)}
+ Q^{(1)} e_2^{(2)} R^{(1)} P^{(1)}
+ Q^{(1)} e_2^{(2)} R^{(1)} Q^{(1)} R'^{(1)}. \\
&=
P^{(1)} + Q^{(1)} e_2^{(2)} R^{(1)} R'^{(1)}.
\end{align*}

We refer to this as \textit{idempotent pruning}, as orthogonal idempotents on the control site annihilate cross-terms, preventing the combinatorial branching that arises under naive composition. This behavior has a direct computational meaning: the idempotents $P^{(1)}$ and $Q^{(1)}$ encode a binary sector on the control qubit, so operators are represented by a fixed number of sector branches under composition, rather than growing combinatorially as in general multivector expansions.

\begin{lemma}
Any word
$
W \in \langle T_1, \mathrm{CNOT}_{1\to 2} \rangle
$
admits a sector--rotor normal form
\[
W = P^{(1)} + Q^{(1)} K,
\qquad K \in \mathcal{C}\ell_{2,0}^{\widehat{\otimes}2},
\]
which is preserved under composition.
\end{lemma}

\begin{proof}
We use induction on the word length in the monoid generated by $\{T_1,\mathrm{CNOT}_{1\to 2}\}$. As shown above, the statement holds for generators, with $K = R^{(1)}$ for $T_1$ and $K = e_2^{(2)}$ for $\mathrm{CNOT}_{1\to 2}$. Assume $W$ has this sector--rotor form. It suffices to show that right multiplication by either generator preserves this form.

\smallskip
\noindent
(i) Composition with $T_1$:
\begin{align*}
W T_1
&=
\big(P^{(1)} + Q^{(1)} K\big)
\big(P^{(1)} + Q^{(1)} R^{(1)}\big) \\
&=
P^{(1)}P^{(1)}
+ P^{(1)} Q^{(1)} R^{(1)}
+ Q^{(1)} K P^{(1)}
+ Q^{(1)} K Q^{(1)} R^{(1)}. \\
&=
P^{(1)} + Q^{(1)} K Q^{(1)} R^{(1)}.
\end{align*}
By idempotent relations on the control site, $W T_1$ is of the form $P^{(1)} + Q^{(1)} K'$ with
\(
K' := K Q^{(1)} R^{(1)} \in \mathcal{C}\ell_{2,0}^{\widehat{\otimes}2}.
\)

\smallskip
\noindent
(ii) Composition with $\mathrm{CNOT}_{1\to 2}$:
\begin{align*}
W\,\mathrm{CNOT}_{1\to 2}
&=
\big(P^{(1)} + Q^{(1)} K\big)
\big(P^{(1)} + Q^{(1)} e_2^{(2)}\big) \\
&=
P^{(1)}P^{(1)}
+ P^{(1)} Q^{(1)} e_2^{(2)}
+ Q^{(1)} K P^{(1)}
+ Q^{(1)} K Q^{(1)} e_2^{(2)}\\
&=
P^{(1)} + Q^{(1)} K Q^{(1)} e_2^{(2)},
\end{align*}
which is again of sector--rotor form with
\(
K' := K Q^{(1)} e_2^{(2)}.
\)
By induction, any word in $\{T_1,\mathrm{CNOT}_{1\to 2}\}$ has a $P, QK$ decomposition.
\end{proof}

\begin{comment}
This reflects a geometric property of how left and right actions interact with the Peirce decomposition of $\widetilde{\mathcal S}$. Operators in higher levels correspond to attaching right-phase transformations to progressively finer idempotent resolutions of the spinor space. A full analysis of multi-qubit Peirce refinement and its implications for higher hierarchy levels is deferred to future work.
\end{comment}

\subsubsection{Generalized Semi-Clifford Gates as a Monomial Factorization}

A unitary operator $U$ is said to be \emph{generalized semi-Clifford} if it admits a decomposition
$
U = C_1 \, \pi \, D \, C_2,
$
where $C_1, C_2 \in \mathcal{C}_2$, $\pi$ is a permutation of the computational basis, and $D$ is diagonal in that basis. Beigi and Shor \cite{Beigi2009C3} showed that every operator in the third level $\mathcal{C}_3$ of the Clifford hierarchy has such a decomposition, which also has algebraic interpretation in terms of the Peirce decomposition of the spinor space. For each bitstring $x=(x_1,\dots,x_n)\in\{0,1\}^n$, define the idempotent
\[
\Pi_x
=
\prod_{i=1}^n
\bigl(P^{(i)}\bigr)^{1-x_i}
\bigl(Q^{(i)}\bigr)^{x_i}.
\]
Then
$
\Pi_x^2 = \Pi_x, \;\
\Pi_x \Pi_y = 0 \ (x \neq y), \;\
\sum_x \Pi_x = 1,
$
so that the identity decomposes into a complete family of mutually orthogonal sector projectors.

Now, consider the subgroup
$
\Delta
=
\left\{
\sum_x R_x \, \Pi_x \;:\; R_x = e^{J\theta_x}
\right\},
$
consisting of operators that act diagonally by right rotors on each sector. Any permutation $f$ of the index set induces an operator $P_f$ that relabels these sectors by
$
P_f \, \Pi_x \, P_f^{-1} = \Pi_{f(x)}.
$
Together, these generate the \emph{monomial subgroup}
$
\mathcal{M}
=
\Delta \rtimes S_{2^n},
$
whose elements permute the sector idempotents $\{\Pi_x\}$ and assign a phase factor to each sector. The Beigi--Shor theorem becomes
\[
\forall U \in \mathcal{C}_3
\quad \Longrightarrow \quad
\exists\, C_1, C_2 \in \mathcal{C}_2
\ \text{such that}\
C_1^{-1} U C_2^{-1} \in \mathcal{M}.
\]
Thus every $\mathcal{C}_3$ operator is Clifford-equivalent to a monomial operator relative to the chosen Peirce decomposition, and
\begin{equation}
\mathcal{C}_3 \subseteq \mathcal{C}_2 \, \mathcal{M} \, \mathcal{C}_2.
\end{equation}
In other words, every $\mathcal{C}_3$ operator lies in a double coset of the monomial subgroup with respect to $\mathcal{C}_2$.

This parallels known decompositions of the quantum Clifford group. Bravyi and Maslov \cite{bravyi2021hadamard} showed that any $U \in \mathcal{C}_2$  can be factored as
$
F_1 \, H \, \pi \, F_2,
$
where $H$ is a layer of Hadamard gates, $\pi$ is a permutation, and $F_i$ belong to Hadamard-free Clifford subgroups. 
As a level-$2$ analogue of the generalized semi-Clifford, Clifford operators already decompose into permutation and diagonal-type components relative to a chosen basis. The Beigi--Shor theorem extends this picture to $\mathcal{C}_3$, where, up to Clifford conjugation, operators reduce to a monomial normalizer associated with a finer sector decomposition.

The monomial subgroup is the normalizer of the diagonal subalgebra $\Delta$,
\[
\mathcal{M} = N_G(\Delta), \qquad \mathcal{M}/\Delta \cong S_{2^n}.
\]
The subgroup $\Delta$ plays the role of a torus-like diagonal subalgebra, and permutations of the sector idempotents act as the discrete symmetry group of the Peirce decomposition. This mirrors Bruhat-type decompositions in Lie theory, where a reductive group $G$ can be factored as $G = \bigsqcup_{w \in W} B w B$ in terms of a Borel subgroup $B$ and Weyl group $W$. In the present setting, the quantum Clifford group $\mathcal{C}_2$ plays the role of a frame-changing subgroup, while the monomial subgroup $\mathcal{M}$ provides the discrete symmetry through permutations of the Peirce sectors. In that light, the Clifford hierarchy shows a symmetry-based interpretation; the quantum Clifford group arises as the normalizer of the Pauli subgroup,
$
\mathcal{C}_2 = N_G(\mathcal{P}_n),
$
while the next level reduces, up to Clifford conjugation, to the normalizer of a finer commutative subalgebra determined by the Peirce decomposition.

\begin{comment}
This suggests that higher levels of the Clifford hierarchy may correspond to normalizers of progressively refined commutative algebras, yielding an algebraic interpretation of the hierarchy in terms of nested symmetry reductions.
\end{comment}
\section{Conclusion and Outlook}

We have developed a real formulation of $n$-qubit quantum information based on the tensor product Clifford algebra $\mathcal{C}\ell_{2,0}(\mathbb{R})^{\otimes n}$, in which states and operators arise within a unified algebraic framework. Spinor states appear as elements of a minimal left ideal, while gates are endomorphisms of its $J$-closure. The bivector $J = e_{12}$ provides the intrinsic generator of complex structure through right multiplication, without the need for an external complex field while preserving standard unitary dynamics.
The resulting State–Operator Clifford Compatibility expresses an equivariance between geometric multiplication and state evolution, allowing stabilizer dynamics and Clifford operations to be realized through local geometric products. The role of “Clifford” is restored to its geometric origin—quantum evolution is governed by Clifford conjugation, and the Gottesman--Knill theorem follows directly from the locality of geometric multiplication in the algebra.

The Peirce decomposition reveals a sector structure on the spinor space, in which idempotents define orthogonal components and operator action is constrained by compatibility with the physical ideal. Within this setting, the generalized semi-Clifford theorem becomes transparent: every $\mathcal{C}_3$ operator is Clifford-equivalent to a monomial transformation acting on these sectors. This shifts the emphasis from recursive definitions to the interaction between sector structure, symmetry, and diagonalization, indicating that higher levels of the hierarchy reflect controlled transitions between distinct algebraic splittings. A central open problem is to determine how these sector transitions compose and whether they admit compressed or canonical representations that improve on existing simulation paradigms. 

Additionally, the elementary data of topological quantum computation admits a natural representation within this framework at the level of the three-anyon fusion space. In the spinor basis $\{P, e_2 P\}$, the Fibonacci recoupling transformation is realized by left multiplication with 
$
F = \varphi^{-1} e_1 + \varphi^{-1/2} e_2,
$
while braiding is a sector-controlled right rotor
$
R = P e^{-4\pi J/5} + Q e^{3\pi J/5}.
$
Extending this beyond the three-anyon setting, and determining whether higher fusion constraints can be captured through compatible idempotent structures, remains an open problem.

Beyond quantum circuit analysis, this framework connects to symmetry-driven learning architectures. Clifford-equivariant neural networks and Clifford-steerable convolutional models \cite{ruhe2023,weiler2024cliffordsteerable} demonstrate that geometric algebra provides a natural language for encoding equivariance. The sector decomposition developed here provides a complementary viewpoint in which transformations act by permuting and phasing algebraic sectors, yielding architectures that operate on Clifford representations rather than raw amplitudes. Finally, the sector structure supports alternative computational models in which quantum evolution is expressed through transformations between algebraic sectors instead of full amplitude vectors. Operations that preserve or refine this decomposition admit sparse or locality-preserving realizations, with potential advantages for simulation, compilation, and representation. A systematic development of these computational consequences—including idempotent-based encodings, sectorwise update rules, and their algorithmic complexity—will be developed in subsequent work.

\appendix
\section{Grover's Search}

To illustrate the formalism, we show that amplitude amplification can be expressed naturally in the idempotent/ideal representation, where reflection-based quantum algorithms appear as products of idempotent involutions. We work out the derivation for $n=2$ and marked string $w=11$, where a single Grover step maps the uniform state $\psi_s$ to the marked basis state $\psi_{11}$. The computation relies on the following Clifford multiplication identities
\begin{equation}
Qe_2P=e_2P,\qquad e_2Pe_2=Q.
\end{equation}
\noindent
\paragraph{\textit{Step 1. Two-qubit basis and the uniform state.}}
Write $P_2 = P^{\otimes2}=P^{(1)}P^{(2)}$ (tensor product of commuting idempotents on distinct sites). The four ideal basis elements are
\[
\psi_{00} = P^{(1)} P^{(2)},\quad
\psi_{01}=P^{(1)} (e_2^{(2)}P^{(2)}),\quad
\psi_{10}=(e_2^{(1)} P^{(1)}) P^{(2)},\quad
\psi_{11}=(e_2^{(1)}P^{(1)}) (e_2^{(2)}P^{(2)}).
\]

With our Hadamard element on a single site as
$
h:=\frac{e_1+e_2}{\sqrt2}\in \mathcal{C}\ell_{2,0}(\mathbb R),
$
we define the uniform superposition state in the ideal by
$\psi_s := h^{\otimes2}P^{\otimes2}\in \mathcal S_2.$
Now compute $h^{(i)}P^{(i)}$ on one site,
\[
h^{(i)}P^{(i)}
=\frac{e_1^{(i)}+e_2^{(i)}}{\sqrt2}P^{(i)}
=\frac{e_1^{(i)}P^{(i)}+e_2^{(i)}P^{(i)}}{\sqrt2}
=\frac{P^{(i)}+e_2^{(i)}P^{(i)}}{\sqrt2},
\]
since $e_1^{(i)}P^{(i)}=P^{(i)}$. Therefore by distributivity,
\begin{align*}
\psi_s
= h^{\otimes 2}P^{\otimes 2}
= (h^{(1)}P^{(1)})(h^{(2)}P^{(2)})
&= \frac{(P^{(1)}+e_2^{(1)}P^{(1)}) (P^{(2)}+e_2^{(2)}P^{(2)})}{2} \\
&= \frac{P^{(1)}P^{(2)}+P^{(1)}e_2^{(2)}P^{(2)}+e_2^{(1)}P^{(1)} P^{(2)}+e_2^{(1)}P^{(1)}e_2^{(2)}P^{(2)}}{2} \\
&= \frac{\psi_{00}+\psi_{01}+\psi_{10}+\psi_{11}}{2}.
\end{align*}

\noindent
\paragraph{\textit{Step 2. The oracle involution flip.}}
For the marked string $w=11$ we have $\psi_{11}=e_2^{(1)}e_2^{(2)}P_2$. By reversion, which reverses order and fixes $P_2$, $\widetilde{\psi_{11}} = P_2 e_2^{(1)} e_2^{(2)}$. The marked projector is
$
\Pi_w := \psi_{11}\widetilde{\psi_{11}}.
$
Now,
\[
\Pi_w 
= (e_2^{(1)}e_2^{(2)}P_2)(P_2e_2^{(1)}e_2^{(2)})
= (e_2^{(1)}P^{(1)}e_2^{(1)})(e_2^{(2)}P^{(2)}e_2^{(2)})
= Q^{(1)}Q^{(2)}
\]
by the single-qubit identity $e_2Pe_2=Q$ applied at each site. The oracle involution is
$
R_w = 1-2Q^{(1)}Q^{(2)}.
$
Now observe,
\[
Q^{(1)}Q^{(2)}\,\psi_{11}
= (Q^{(1)} e_2^{(1)}P^{(1)})(Q^{(2)} e_2^{(2)}P^{(2)})
= (e_2^{(1)}P^{(1)})(e_2^{(2)}P^{(2)})
= \psi_{11},
\]
since $Qe_2P=e_2P$ on each site, while for any basis element with at least one $0$ bit,
$Q^{(i)}P^{(i)}=0$ implies $Q^{(1)}Q^{(2)}$ annihilates it. Therefore
$
Q^{(1)}Q^{(2)}\,\psi_s = \frac{1}{2}\psi_{11},
$
and so
\[
R_w\psi_s
= \Big(1-2Q^{(1)}Q^{(2)}\Big)\psi_s
= \psi_s - \psi_{11}
= \frac{\psi_{00}+\psi_{01}+\psi_{10}-\psi_{11}}{2},
\]
which is exactly the usual phase flip on the marked component, obtained entirely by Clifford multiplication.

\paragraph{\textit{Step 3. The diffusion involution.}}
Define the Clifford analogue of $|s\rangle\langle s|$ as the rank-one diffusion projector
\[
\Pi_s := \psi_{s}\widetilde{\psi_s}
= (h^{\otimes 2} P^{\otimes 2})(P^{\otimes 2} h^{\otimes 2})
= h^{\otimes 2} P^{\otimes 2} h^{\otimes 2}.
\]
Using that $h^{(i)}e_1^{(i)}h^{(i)}=e_2^{(i)}$ (Hadamard swaps $\sigma_z \leftrightarrow \sigma_x$ by conjugation),
we obtain
\[
h^{(i)}P^{(i)}h^{(i)}
= \frac{1+h^{(i)}e_1^{(i)}h^{(i)}}{2}
= \frac{1+e_2^{(i)}}{2}
=: P_{+}^{(i)},
\]
the projector onto the $+1$ eigenspace of $e_2^{(i)}$.
Hence
$
\Pi_s = \prod_{i=1}^2 P_{+}^{(i)}.
$
Now compute the action of $P_{+}^{(i)}$ on the one-qubit ideal basis,
\[
P_{+}^{(i)}P^{(i)} = \frac{(1+e_2^{(i)})P^{(i)}}{2} = \frac{P^{(i)}+e_2^{(i)}P^{(i)}}{2}, \qquad
P_{+}^{(i)}(e_2^{(i)}P^{(i)})
= \frac{(1+e_2^{(i)})e_2^{(i)}P^{(i)}}{2}
= \frac{P^{(i)}+e_2^{(i)}P^{(i)}}{2}.
\]
For either one-qubit basis state $\psi_{0}$ or $\psi_{1}$ ($x \in \{0,1\}$), one has
$
P_{+}^{(i)}\,\psi_{x}
=
\frac{1}{2}\bigl(P^{(i)} + e_2^{(i)}P^{(i)}\bigr),
$
and for any two-qubit basis state $\psi_{xy}$,
\[
\Pi_s\,\psi_{xy}
=
\frac{\psi_{00}+\psi_{01}+\psi_{10}+\psi_{11}}{4}
=
\frac{1}{2}\psi_s,
\]
meaning $\Pi_s$ projects every basis vector onto $\frac12\psi_s$.  For the oracle state
$
R_w\psi_s
$
the coefficients sum to $1$, so
$
\Pi_s(R_w\psi_s)=\frac{1}{2}\psi_s.
$
Finally, define the diffusion involution $
R_s := 2\Pi_s - 1
     = 2 h^{\otimes 2} P^{\otimes 2} h^{\otimes 2} - 1,
$
which under $\rho$ is the usual Hilbert-space diffusion operator
$
H^{\otimes 2}\!\left(2|00\rangle\langle 00| - I\right)H^{\otimes 2}.
$ 
Then
\[
R_s(R_w\psi_s) =  (2\Pi_s - 1) (R_w\psi_s) = 2\Pi_s(R_w\psi_s) - (R_w\psi_s) =
2\cdot\frac{1}{2}\psi_s-(R_w\psi_s)
=
\psi_s-\frac{\psi_{00}+\psi_{01}+\psi_{10}-\psi_{11}}{2}
=
\psi_{11}.
\]
Thus one Grover step yields the marked state,
$
G\psi_s = (R_s R_w)\psi_s = \psi_{11}.
$

\bibliographystyle{plain}
\bibliography{references}

\end{document}